\documentclass[3p,times]{elsarticle}

\usepackage{ecrc}
\usepackage{color}

\volume{00}

\firstpage{1}

\journalname{Physics Procedia}

\runauth{}

\jid{phpro}

\jnltitlelogo{Physics Procedia}

\usepackage{amssymb}

\usepackage[figuresright]{rotating}

\begin{document}

\begin{frontmatter}



\dochead{}

\title{{\small12$^{\rm th}$ International Conference on Muon Spin Rotation, Relaxation and Resonance}\\
Microscopic Magnetic Nature of the\\Quasi-one-Dimensional Antiferromagnet BaCo$_{2}$V$_{2}$O$_{8}$}


\author[ETH,PSI]{M.~M\aa{}nsson\corref{cor1}},
\cortext[cor1]{Corresponding author. Tel.: +41-(0)56-310-5534 ; fax: +41-(0)44-633-1282}
\ead{mansson@phys.ethz.ch}
\author[ETH,PSI]{K.~Pr$\check{\rm s}$a}
\author[TCRDL]{J.~Sugiyama}
\author[TCRDL]{H.~Nozaki}
\author[LMU]{A.~Amato}
\author[Osaka]{K.~Omura}
\author[IMR]{S.~Kimura}
\author[Osaka]{and M.~Hagiwara}

\address[ETH]{Laboratory for Solid state physics, ETH Z\"{u}rich, CH-8093 Z\"{u}rich, Switzerland}
\address[PSI]{Laboratory for Neutron Scattering, Paul Scherrer Institute, CH-5232 Villigen PSI, Switzerland}
\address[TCRDL]{Toyota Central Research and Development Labs. Inc., Nagakute, Aichi 480-1192, Japan}
\address[LMU]{Laboratory for Muon-Spin Spectroscopy, Paul Scherrer Institut, CH-5232 Villigen PSI, Switzerland}
\address[Osaka]{KYOKUGEN, Osaka University, Machikaneyama 1-3, Toyanaka 560-8531, Japan}
\address[IMR]{Institute for Materials Research, Tohoku University, Katahira 2-1-1, Sendai 980-8577, Japan}

\begin{abstract}
The title compound belongs to a wide group of quasi-one-dimensional (Q1D) antiferromagnets (AF) and its Co$^{2+}$ ions form Q1D screw-chains along the $c$-axis. We here present the first investigation of the microscopic magnetic nature of a single crystalline BaCo$_2$V$_2$O$_8$ sample using $\mu^{+}$SR. Our data reveal the presence of several clear muon frequencies below $T_N$ indicating the onset of a long-range order. Above 5 K, the $\mu^+$SR spectra is well fitted to a simple power-exponential relaxing function. The temperature dependence of the relaxation-rate ($\lambda$) as well as the power ($n$) display a clear anomaly around $T = 44$ K, indicating the onset of short-range 1D correlations.
\end{abstract}


\end{frontmatter}


\section{Introduction}
The title compound, BaCo$_{2}$V$_{2}$O$_{8}$, belongs to a wide group of so-called quasi-one-dimensional (Q1D) antiferromagnets (AF). In more detail, it belongs to the 1D spin chain systems with spin-3/2 and the crystal structure \cite{He1} has tetragonal symmetry of space group $I41/acd$ with $a=12.444(1)~${\AA}, $c=8.415(3)~${\AA}, and $Z=8$. All magnetic Co$^{2+}$ ions are equivalent and build up edge-sharing CoO$_6$ octahedra that in turn form screw-chains along the $c$-axis [see \textbf{1-4} in Fig.~1(a-b)]. The screw-chains are separated by nonmagnetic VO$_4$ (V$^{5+}$) tetrahedra as well as the Ba$^{2+}$ ions, resulting in a Q1D structure.
The physics in this group of materials is governed by a strong spin-spin coupling along the 1D direction, together with a much weaker coupling along the other directions \cite{Sugiyama_Q1D_PRL}. Further, it is well known that an ideal 1D AF spin system does not show long-range ordering (LRO) above $T=0~$K due to strong quantum spin fluctuations, causing such unconventional phenomena as quantum phase transitions (QPT). One of the most popular way to study QPT is to induce a magnetic order in a spin gap system by applying an external magnetic field \cite{Oosawa,Nikuni}. However, there exist some special cases, including BaCo$_{2}$V$_{2}$O$_{8}$, where the LRO is instead destroyed by the applied field \cite{He2}.
\begin{figure}[t]
  \begin{center}
    \includegraphics[keepaspectratio=true,width=140 mm]{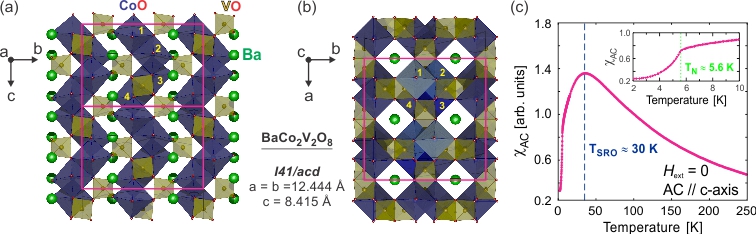}
  \end{center}
  \caption{Crystal structure of BaCo$_{2}$V$_{2}$O$_{8}$ seen along the (a) a-axis and (b) c-axis, showing the quasi-1D screw-chains created by edge-shared CoO$_6$ octahedra (1..4) that run along the c-axis. (c) AC susceptibility [$\chi_{\rm AC}(T)$] with the onset of possible short-range order at $T_{\rm SRO}=30$~K and (inset) long-range AF order at $T_{\rm N}^{\chi}=5.6$~K.
  }
  \label{fig:structure}
\end{figure}
The temperature dependence of the magnetic susceptibility in our sample [Fig.~1(c)] displays a broad peak around 30~K, typical for 1D short-range ordering (SRO). The susceptibility then decreases with decreasing temperature until a sudden cusp appears at $T_{\rm N}^{\chi}\approx5.6$~K [inset of Fig.~1(c)]. By fitting to a Curie-Wiess law, a negative Weiss temperature ($\Theta_{\rm CW}=-53.3$~K) is obtained, suggesting AF (intrachain) interactions. From anisotropic susceptibility measurements \cite{He1} it was evident that the easy magnetization is along the $c$-axis. Also, when the magnetic field is perpendicular the $c$-axis, a sharp peak is visible at the 5~K transition, indicating the onset of a clear 3D AF LRO. Such statement found further supports in the heat capacity data with $H=0~$T. Here no anomaly is found around 30~K, while a very sharp peak is visible at 5~K. The data also indicate that a large portion of the magnetic entropy is bound up in short-range magnetic correlations, extending up to temperatures well above 5~K. In more detailed field-dependent measurements of the susceptibility and heat capacity \cite{He2}, a field induced order-disorder transition was revealed. Field dependent magnetic susceptibility \cite{He2} showed how the 5~K transition shifts toward lower temperatures with increasing magnetic field and finally approaches 0~K for H$_{\rm c}$~=~3.9~T.
In very recent high-field heat capacity \cite{Kimura2} and neutron diffraction \cite{Kimura3} experiments, a novel type of field induced magnetic order was found for T~$\leq$~1.8~K and $H_{\rm c}\geq3.9$~T. This new phase was determined to be an incommensurate spin structure caused by quantum fluctuations, fitting well to theoretical predictions for a so-called Tomonaga-Luttinger liquid (TLL). Hence, this compound was revealed to be a good realization of quasi-1D S = 1/2 Ising-like AF with the transition field $H_{\rm c}=3.9$~T \cite{Kimura1,He2}

\section{\label{sec:E}Experimental Details}
A single piece of BaCo$_{2}$V$_{2}$O$_{8}$ single crystal (${\O}$ = 6 mm, 1 mm thick) was attached to a low-background (fork-type) sample holder using very thin Al-coated Mylar tape. In order to make certain that the muon stopped primarily inside the sample, we ensured that the side facing the muon beamline was only covered by a single layer of mylar tape. Subsequently, $\mu^+$SR spectra were measured at the Swiss Muon Source (S$\mu$S), Paul Scherrer Institut, Villigen, Switzerland. Zero-field (ZF) and weak transverse-field (wTF) spectra were collected at the General Purpose Spectrometer (GPS). Data was collected for both non-spin-rotated mode (NSR) and spin-rotated (SR) mode. The experimental setup and techniques are described in detail elsewhere \cite{Kalvius}.

\section{\label{sec:E}Results \& Discussion}
\begin{figure}[t]
  \begin{center}
    \includegraphics[keepaspectratio=true,width=150 mm]{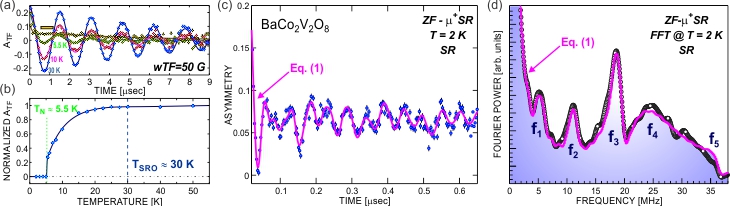}
  \end{center}
  \caption{(a) wTF~=~50~G time spectra at different temperatures fitted to a combination of a relaxing cosine function and a simple exponentially relaxing signal. (b) $T$--dependence the wTF asymmetry ($A_{\rm TF}$). (c) ZF time spectrum at $T=2$~K acquired in spin-rotated (SR) mode. (d) Fast Fourier transform (FFT) of the time spectrum in (c) showing the 5 different frequencies $f_{i}$ ($i=1..5$). Solid magenta line in (c-d) are fits to Eq.~(1).
    }
  \label{fig:structure}
\end{figure}
First, weak transverse-field (wTF = 50 G) $\mu^{+}$SR time spectra were acquired [see Fig.~2(a)] as a function of temperature. In Fig.~2(b) the $T$-dependence of the wTF asymmetry ($A_{\rm TF}$) is shown. Upon heating, at the known bulk magnetic transition $T_{\rm N}^{\rm wTF}\approx5.3$~K, $A_{\rm TF}$ display a sudden increase but only recovers approximately 25\% of its maximum value. Upon further heating $A_{\rm TF}$ then increases more slowly until the full $A_{\rm TF}$ is recovered around $T=30$~K, which fits well to the maximum of the $\chi_{\rm AC}(T)$ curve. To further investigate the details of this $T$--range, more sensitive zero-field (ZF) measurements were performed. At lowest temperature, $T=2$~K a static magnetic order is revealed from the clear presence of spontaneous muon precession [Fig.~2(c)]. The ZF data was well fitted by the combination of five damped cosine oscillations and a \textit{tail} signal due to the field component parallel to the initial muon spin:
\begin{eqnarray}
 A_0 \, P_{\rm ZF}(t) = A_{\rm tail}~e^{-\lambda_{\rm tail} t} + \sum^{5}_{i=1}A_{\rm i}~e^{-\lambda_{\rm i} t}\cos(2\pi f_{\rm i}\cdot{}t+\phi_i)~,
\label{eq:ZFfit}
\end{eqnarray}
The obtained frequencies and the resulting field distribution map agrees well with the fast Fourier transform (FFT) of the $\mu^+$SR time spectra [Fig.~2(d)]. These results compare well to previous data from powder samples \cite{Kawasaki}. However, it should be noted that the quality of our single crystal data is much higher. As a result, among several things, we additionally managed to resolve the higher $f_{5}$ component, which makes the fit to the data significantly more accurate (especially obvious in the FFT domain). In order to further investigate the formation of static magnetic order, ZF-$\mu^+$SR spectra were collected for 2~K~$\leq T<$~150~K. It is found that as $T$ increases the oscillation frequencies are gradually slowing down and finally disappears around $T_{\rm N}^{\rm ZF}=5.35$~K. From fits to Eq.~(1) it is possible to extract the magnetic order parameter as shown in Fig.~3(a), which was found to be well fitted to the commonly used phenomenological formula:
\begin{eqnarray}
 f_{i}(T) = f_{\rm T\rightarrow 0 K} \cdot \left[1-\left(\frac{T}{T_{\rm N}}\right)^{\alpha}\right]^{~\beta},
\label{eq:ZFfit}
\end{eqnarray}
where $\beta$ is a parameter describing the dimensionality of the magnetic order \cite{Collins}. Using $T_{\rm N}=5.35$~K the fits yield only slightly different values for $f_{1-4}$ and the average value obtained is $\beta=0.32$. This values almost perfectly agrees with predictions for a 3D Ising system \cite{Collins}. Noteworthy is also that the high $f_{5}$ suddenly disappears well before $T_{\rm N}$ is reached. This fact is hard to explain but one could consider a connection to the low-temperature field-induced TLL phase that emerges above $H_{\rm c}$~=~3.9~T in this particular temperature range \cite{Kimura3}. Further investigations under magnetic field are however needed and underway to clarify this matter.

Above $T_{\rm N}$ the ZF time spectra display no oscillations but rather an exponential decay, indicating the presence of dynamic spin correlations. Accordingly, the data is well fitted to a power-exponential \cite{Phillips} relaxing function:
\begin{eqnarray}
 A_0 \, P_{\rm ZF}(t) = A\cdot exp\left[(-\lambda t)^{n}\right]~,
\label{eq:KT}
\end{eqnarray}
The temperature dependence of the relaxation-rate ($\lambda$) displays a typical critical behavior [Fig.~3(b)] in the vicinity of $T_{\rm N}$ as the dynamics of the Co spins are gradually slowing down. In addition, around $T_{SR}\approx44$~K, the $\lambda(T)$ curve displays an anomaly possibly revealing the onset of short-range spin correlations. Also the critical exponent ($n$) displays an intriguing temperature evolution below $T_{SR}\approx44$~K. Coming from higher temperatures, $n$ corresponds to an isotropic system ($n\approx1.5$) but then suddenly decreases below $n=1$ for $T<T_{SR}$ indicating the evolution of anisotropic spin dynamics. In the close vicinity of $T_{\rm N}$ the power reaches $n=1/3$, which corresponds to a purely 1D case. At $T_{\rm N}$ the system then enters into a static 3D (!) magnetic order, as shown above. Finally, it is worth noticing that even at $T=150$~K, the BaCo$_2$V$_2$O$_8$ compound does not display a Kubo-Toyabe behavior, indicating the persistence of isotropic dynamic spin correlations well above the transition temperature(s).
\begin{figure}[t]
  \begin{center}
    \includegraphics[keepaspectratio=true,width=150 mm]{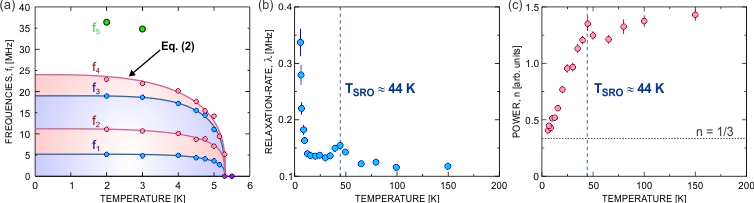}
  \end{center}
  \caption{(a) Magnetic order parameter showing the temperature dependence of the muon precession frequencies $f_{i}$ ($i=1..5$). Solid lines are fits to Eq.~2. Above $T_{\rm N}$ the $\mu^{+}$SR time spectra are well fitted by a power-exponential relaxation according to Eq.~(3). (b-c) Display the temperature dependence of the relaxation-rate ($\lambda$) and the critical exponent/power ($n$), respectively.
    }
  \label{fig:structure}
\end{figure}
\section{\label{sec:E}Summary}
In summary, we have presented the first $\mu^+$SR investigation of a single crystalline sample of the Q1D compound BaCo$_2$V$_2$O$_8$. Going from hight to low temperature the compound shows already at 150 K the presence of isotropic spin correlations and the ZF data is fitted to a stretched exponential. At $T_{SR}\approx44$~K, $\lambda$ displays an anomaly and at the same point the power ($n$) starts to decrease, indicating the onset of anisotropic correlations. Decreasing the temperature further, $\lambda$ displays a strong increase i.e. a critical slowing down of the Co moments in the vicinity of $T_{\rm N}$. In the same region, $n$ obtains a value close to 1/3, indicating that the spin correlations have a distinct 1D character. At $T_{\rm N}^{\rm ZF}=5.35$~K the compound finally enters into a static AF order and from fits to the magnetic order parameter it is evident that the character changes to a 3D (Ising) character. This sudden change of dimensionality can be understood from the fact that one--dimensional chains cannot display a stable long--range AF order for $T>0~$K \cite{Landau,MerminWagner}. As long as the spins are dynamic, the 1D character is sustained. However, as soon as the fluctuations slow down and the system becomes static, the spins are locked into a 3D-order by (weak) inter-chain interactions.

\paragraph{\textbf{Acknowledgments}}\

This work was performed using the \textbf{GPS} muon spectrometer at the Swiss Muon Source (S$\mu$S) of the Paul Scherrer Institut (PSI), Villigen, Switzerland and we are thankful to the instrument staff for their support. All the $\mu^{+}$SR data was fitted using \texttt{musrfit} \cite{musrfit} and the images involving crystal structure were made using the DIAMOND software. This research was financially supported by the Swiss National Science Foundation (through Project 6, NCCR MaNEP) and Toyota Central Research \& Development Labs. Inc. and a Grant-in-Aid for Scientific Research B (No. 20340089) from the MEXT, Japan.

\paragraph{\textbf{References}}\

\label{}





\bibliographystyle{elsarticle-num}



\end{document}